\documentclass[12pt]{iopart}
\usepackage{iopams}  
\usepackage{graphicx}
\def\d{{\mathrm{d}}}
\def\dbar{{\mathchar'26\mkern-12mu \d}} 

\def\x{{\mathbf{x}}}
\def\n{{\mathbf{n}}}

\def\O{{\mathcal{O}}}
\def\A{{\mathcal{A}}}
\def\B{{\mathcal{B}}}
\def\S{{\mathcal{S}}}

\def\sign{{\mathrm{sign}}}

\begin{document}
\def\TITLE{Clausius entropy for arbitrary bifurcate null surfaces}
\title{\TITLE}
\author{Valentina Baccetti {\rm and} Matt Visser}
\address{School of Mathematics, Statistics, and Operations Research\\
Victoria University of Wellington, PO Box 600 \\
Wellington 6140, New Zealand}
\ead{valentina.baccetti@msor.vuw.ac.nz \textrm{and} matt.visser@msor.vuw.ac.nz}
\begin{abstract}
Jacobson's thermodynamic derivation of the Einstein equations was originally applied only to local Rindler horizons.  But at least some parts of that construction can usefully be extended to give meaningful results for arbitrary bifurcate null surfaces. As presaged in Jacobson's original article, this more general construction sharply brings into focus the questions: 
Is entropy objectively ``real"? Or is entropy in some sense subjective and observer-dependent? These innocent questions open a Pandora's box of often inconclusive debate. A consensus opinion, though certainly not universally held, seems to be that Clausius entropy (thermodynamic entropy, defined via a Clausius relation $\d S = \dbar Q/T$) should be objectively real, but that the ontological status of statistical entropy (Shannon or von Neumann entropy) is much more ambiguous, and much more likely to be observer-dependent. This question is particularly pressing when it comes to understanding Bekenstein entropy (black hole entropy). To perhaps further add to the confusion, we shall argue that even the Clausius entropy can often be observer-dependent.  In the current article we shall conclusively demonstrate that one can meaningfully assign a notion of Clausius entropy to arbitrary bifurcate null surfaces --- effectively defining a ``virtual Clausius entropy'' for arbitrary ``virtual (local) causal horizons''.  As an application, we see that we can implement a version of the generalized second law (GSL) for this virtual Clausius entropy. This version of GSL can be related to certain (nonstandard) integral variants of the null energy condition (NEC). 
Because the concepts involved are rather subtle, we take some effort in being careful and explicit in developing our framework.
In future work we will apply this construction to generalize Jacobson's derivation of the Einstein equations.

\vskip 10 pt
\noindent{\it Keywords}:  Entropy, Clausius entropy, thermodynamic entropy, statistical entropy, Bekenstein entropy, bifurcate null surfaces, causal horizons, generalized second law.
arXiv:1303.3185 [gr-qc]

\vskip 10 pt
\noindent 
13 March 2013; 8 April 2013; 25 September 2013; \LaTeX-ed \today

\end{abstract}

\pacs{89.70.Cf;  89.70.-a;  03.67.-a}

\maketitle

\bigskip
\hrule
\bigskip
\markboth{\TITLE}{}
\tableofcontents
\bigskip
\hrule
\bigskip
\markboth{\TITLE}{}
\clearpage
\markboth{\TITLE}{}
\section{Introduction}\label{S:intro}

Jacobson's thermodynamic derivation of the Einstein equations~\cite{Jacobson} has had, and continues to have, a profound influence on our understanding of the interface between thermodynamics and geometry.  Jacobson's original construction associated heat fluxes and entropies only to local Rindler horizons~\cite{Jacobson}, but left open the question as to whether some suitable notion of entropy could meaningfully be assigned to a broader class of null causal surfaces.    

The construction presented in the current article addresses this point, and  is considerably more general than Jacobson's approach. We shall soon see that while the bifurcate nature of the local Rindler horizon is essential to the construction,  other Rindler-specific features can easily be discarded. In particular, any null surface can be viewed as an observer-dependent causal boundary, a ``virtual'' causal boundary or virtual local horizon --- and our construction can be viewed as providing a notion of virtual entropy for matter crossing \emph{arbitrary} bifurcate virtual causal horizons. 

Ultimately, we will argue that for arbitrary bifurcate null surfaces in curved spacetime, at arbitrary cross-section $\S$ of  the null surface, it is meaningful to define a Clausius entropy ($\,\dbar Q/T$ entropy) in terms of the bifurcation two-surface $\B$, and the affinely parameterized null generators:
\begin{equation}
S_\mathrm{Clausius}(\S) \equiv  S_\B 
+ {2\pi k_B\over\hbar}\; \int_\B^\S \lambda \; T_{ab}\left(x(\xi,\lambda)  \right) \; k_\pm^a k_\pm^b\; \d^2\A \;\d\lambda.
\end{equation}
Since this argued to hold for arbitrary ``virtual" null surfaces, this can be viewed as thereby undermining the ontological reality of entropy; modulo some technical assumptions that we shall be very careful to make explicitly clear. 

Indeed, the ontological status of entropy continues to generate much heated and inconclusive debate. Key questions are: Is entropy objectively ``real''? Or is entropy in some sense subjective and observer-dependent? Part of the issue is that there are many different notions of entropy, and the extent to which they are \emph{universally} equivalent is less than clear. At a minimum, one might wish to consider:
\begin{itemize}
\item Clausius entropy ($\d S = \dbar Q/T$); often called thermodynamic entropy~\cite{Clausius}.
\item Bekenstein entropy; black hole entropy~\cite{Bekenstein}.
\item Statistical entropy; (Shannon~\cite{Shannon}, von Neumann~\cite{vonNeumann}, or entanglement entropy).
\end{itemize}
The extent to which these three notions can \emph{universally} be identified is still a matter of debate, though in certain special cases they can be (and often are) degenerate. 

Other related notions of entropy include Gibbs entropy, Boltzmann entropy, Srednicki entropy, Kolmogorov--Sinai entropy, and 
the Tsallis and Renyi entropies. Multiple attempts have been carried out in order to reconcile these different definitions, see for instance~\cite{Jaynes:1957a, Jaynes:1957b, Bombelli:1986, Srednicki:1993, Coleman:1991, Preskill:1992, Fiola:1994}, and also to separate out and distinguish equilibrium and non-equilibrium notions of entropy~\cite{Langer:1990}. (See also~\cite{Chirco:2009}.) For additional general background see~\cite{Ash:1965, Eisert:2002, Yeung:2002, Jaynes:2003, Watrous:2008, area:2008,  bounds:2009, Renner:2009, area:2010,  Baccetti:2012, Visser:2012}. 

\bigskip
Typically, but not universally, the Clausius entropy is viewed as the most objectively real of these entropies. The Clausius entropy will be the central focus of this article, but even there the situation is extremely subtle. (See for instance the discussion provided by Padmanabhan~\cite{paddy1, paddy2, paddy3}.) These ontological issues are central to Jacobson's ``thermodynamic'' derivation of the Einstein equations~\cite{Jacobson}, where one part of the argument is based on an entanglement entropy interpretation of (a variant of) the Bekenstein entropy, and another part of the argument is based on (a variant of) the Clausius entropy applied to ``local Rindler horizons''.   (These ``local Rindler horizons'' also arise in other situations such as those considered in~\cite{Chirco:2009} and~\cite{Chirco:2010a, Chirco:2010b, Chirco:2011}, and it may prove interesting to see to what extent those constructions could also be generalized.) 
We shall also see that a version of the generalized second law (GSL) can be formulated for this virtual Clausius entropy, and can be related to certain (nonstandard) integral variants of the null energy condition (NEC). 

\section{Strategy}\label{S:strategy}

Instead of addressing Jacobson's ``thermodynamic'' derivation directly we shall in the current article address a more modest goal: To what extent can a Clausius-type notion of entropy be associated with matter crossing \emph{arbitrary} bifurcate null surfaces? 
We shall first work with exact Rindler horizons in flat Minkowski space, systematically and carefully extending the framework until we can successfully deal with arbitrary bifurcate null surfaces in curved spacetime. 

Building on this construction, in future work we plan to more directly address the issue of the extent to which Bekenstein and Clausius entropies can universally be inter-related.
Specifically: For which subset of causal horizons (virtual or otherwise) \emph{should} they be inter-related? Under what situations should these concepts carefully be kept distinct?

\section{Flat Minkowski spacetime}\label{S:flat}

To start the calculation is best to work in flat Minkowski spacetime. That is, for now we are working in the framework of SR, not GR. 

\subsection{Heat flux, temperature, Clausius entropy}\label{SS:fluxes}

To get a handle on the notion of heat flux $\,\dbar Q$ it is convenient to start with a infinitesimal segment of timelike hypersurface, (ruled by a congruence of future-pointing timelike vectors $V^a$, with outward spacelike normal $n^a$, and with hypersurface area element $\d^3\Sigma$), and define a future-pointing flux vector
\begin{equation}
F^a = - T^{ab} \; V_b.
\end{equation}
It is then appropriate to define an infinitesimal heat flux $\;\dbar Q$ by setting
\begin{equation}
\dbar Q =  F^a \; (\d^3 \Sigma)_a  = - T_{ab} \; V^a n^b \; \d^3\Sigma. 
\end{equation}
For finite segments of hypersurface we set
\begin{equation}
\dbar Q =  - \int T_{ab} \; V^a n^b \; \d^3\Sigma. 
\end{equation}
This is our version of Jacobson's equation (1), see reference~\cite{Jacobson}, currently applied to timelike hypersurfaces.
There would be universal agreement that this quantity defines the net energy flux across the segment of timelike hyper-surface, but perhaps less agreement that this energy flux can be equated with a heat flux. (For instance, some authors prefer to identify this quantity with $\d U$, the change in internal energy, while yet others might argue that this quantity should be identified with $\d H$, the change in enthalpy. For current purposes this subtlety is immaterial.) Following Jacobson~\cite{Jacobson}, let us accept the above definition for the sake of argument and see where this identification leads. Note that due to the 
$(-;++ +)$ signature of spacetime this is, perhaps counter-intuitively, the flux of energy in the direction of the normal $n^a$.

We shall now consider a sequence of timelike hypersurfaces, and construct an appropriate null limit. From the way this limit is set up it will soon be clear that we cannot deal with completely arbitrary null surfaces --- the construction intrinsically is set up so that the null limit automatically yields bifurcate null surfaces. 
One of the advantages of Minkowski space is that it is possible to develop some \emph{exact} results, many of which will even hold \emph{globally}. We will subsequently invoke local flatness to extract more limited approximate results in curved spacetimes; approximate results which nevertheless hold up to an explicitly controlled level of accuracy in the vicinity of the bifurcation 2-surface. Ultimately we shall develop a construction valid in arbitrary curved spacetimes.

A key physics step in the computation is to invoke the Unruh effect (acceleration radiation). This (in its original incarnation) is a flat-space SR QFT result whereby an accelerated observer will, (when the QFT is in its usual SR ground state), detect a thermal bath of quantum excitations with a temperature~\cite{Unruh:1976}:
\begin{equation}
k_B T = {\hbar a\over2\pi}.
\end{equation}
We shall use the Unruh effect to \emph{define} the differential Clausius entropy for the matter crossing any timelike hypersurface segment swept out by timelike observers of 4-acceleration $a$ by:
\begin{equation}
\d S = {\dbar Q\over T} = {2\pi k_B \over \hbar a} \;\; \dbar Q 
=  - {2\pi k_B \over \hbar a}  T_{ab} \; V^a n^b \; \d^3\Sigma. 
\end{equation}
For a finite segment of hypersurface we could in principle allow the acceleration $a$ to vary from generator to generator of the timelike hypersurface, (for the time being the acceleration is to be kept constant along each generator, though later on we shall see how to relax this requirement), and would then have
\begin{equation}
\d S =  - {2\pi k_B \over \hbar}  \int {T_{ab}\over a}  \; V^a n^b \; \d^3\Sigma. 
\end{equation}
In all explicit calculations below the hypersurfaces will be set up in such a manner that the acceleration $a$ is a constant over the hypersurface, so that
\begin{equation}
\d S =  - {2\pi k_B \over \hbar a}  \int T_{ab}  \; V^a n^b \; \d^3\Sigma. 
\end{equation}

A tricky point is that the $T_{ab}$ being used here is purely classical, whereas the Unruh temperature is associated with quantum fluctuations in the quantum ground state.  (Jacobson refers to this as considering the  ``thermodynamic limit"~\cite{Jacobson}.) In the presence of excitations above the quantum ground state it can be argued that the Unruh effect provides a lower bound on the physical temperature~\cite{Abreu1, Abreu2, Abreu3, Abreu4}, and so an upper bound on $|\,\dbar Q|/T$.  Furthermore the $\d S$ defined above is a ``virtual'' quantity; there is no actual need for the timelike observers to be objectively real and physically present --- the $\d S$ defined above is what \emph{would be seen} by an imaginary swarm of timelike observers skimming along the timelike hypersurface.

In view of these issues, (identification of the heat flux, identification of the temperature, virtual status of the quantity  $\d S$),  some may refuse to call the quantity $\d S$  a Clausius entropy, and prefer to introduce yet another notion --- perhaps ``Jacobson entropy'' might be appropriate? Be that as it may, provided one accepts this definition, and we hope the reader will agree this is a very plausible and physically interesting object to calculate, most of the technical computations of this article boil down to taking appropriate limits as the acceleration $a$ tends to infinity and the timelike surface becomes null. 

\subsection{Rindler wedges}\label{SS:rindler}

Let us pick an arbitrary spacelike 2-plane in Minkowski space and choose coordinates so that this plane is
\begin{equation}
x^a(x,y) =     (0; \;  x,y, 0).
\end{equation}
Now add past and future light sheets, for convenience in the $+z$ direction. The resulting bifurcate null surface is
\begin{equation}
x^a(t,x,y) =     (t; \;  x,y,  |t|).
\end{equation}
The two null 3-d half-planes are joined by the spacelike bifurcation 2-plane at $t=0$. 
Now pick a sheet of hyperbolic timelike observers ``close'' to that null surface:
\begin{equation}
x^a(\tau; x,y)  = \left( {1\over a}\sinh(a\tau); \; x,y,  {1\over a}  \cosh(a\tau) \right).
\end{equation}
(Eventually we will want to take $a\to\infty$.)
These observers have  4-velocity
\begin{equation}
V^a(\tau; x,y) = \left( \cosh(a\tau); \; 0,0,  \sinh(a\tau) \right); \qquad ||V|| = 1;
\end{equation}
and 4-acceleration
\begin{equation}
A^a(\tau; x,y) =  a \left( \sinh(a\tau); \; 0,0,  \cosh(a\tau) \right); \qquad ||A|| = a;
\end{equation}
while the hyperbolic timelike sheet they sweep out has 4-normal
\begin{equation}
n^a(\tau; x,y) = - \left( \sinh(a\tau); \; 0,0, \cosh(a\tau) \right); \qquad ||n|| = 1.
\end{equation}
Here we have chosen the 4-normal to point towards the Rindler horizon; that is, away from the ``observable'' region containing the virtual timelike observers.
\begin{figure}[!htbp]
\label{F:tangent}
\begin{center}
\includegraphics[scale=0.60, trim = 50 50 50 50]{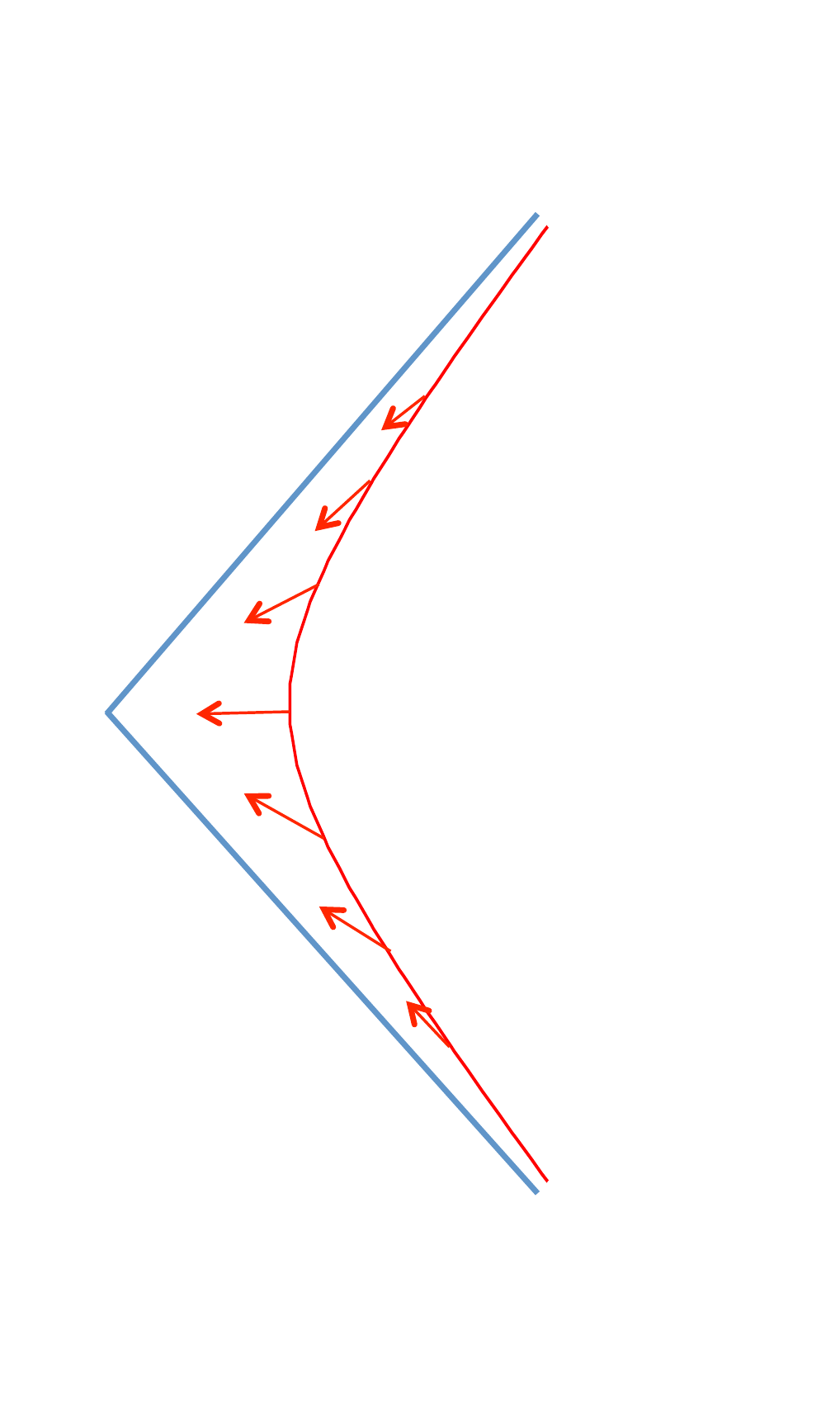}
\end{center}
\caption[Rindler wedge with virtual hyperbolic timelike observer and 4-normal.]{Rindler wedge with virtual hyperbolic timelike observer and 4-normals. \\Note that the 4-normals point towards the Rindler horizon and asymptote to minus the 4-tangent at extremely late and 
extremely early proper times.}

\end{figure}

Note that on the time-like shell we have the restriction
\begin{equation}
T_{ab}(x^a) \to T_{ab}(\tau,x,y).
\end{equation}
Then, setting $\d^2\A = \d x \,\d y$ we have 
\begin{equation}
\dbar Q =  -\int T_{ab} \; V^a n^b  \; \d\tau \; \d^2 \A,
\end{equation}
and so
\begin{equation}
{\dbar Q\over\d\tau} =  -\int T_{ab} \; V^a n^b \;  \d^2 \A,
\end{equation}
whence
\begin{equation}
{\dbar Q\over\d t} =  -\int T_{ab} \; V^a n^b  \; {\d\tau\over\d t}\; \d^2 \A.  
\end{equation}
With current conventions this is the flux of matter crossing the time-like shell in the direction of the Rindler horizon.
\begin{figure}[!htbp]
\label{F:Rindler-flux-0}
\begin{center}
\includegraphics[scale=0.60, trim = 50 50 50 50]{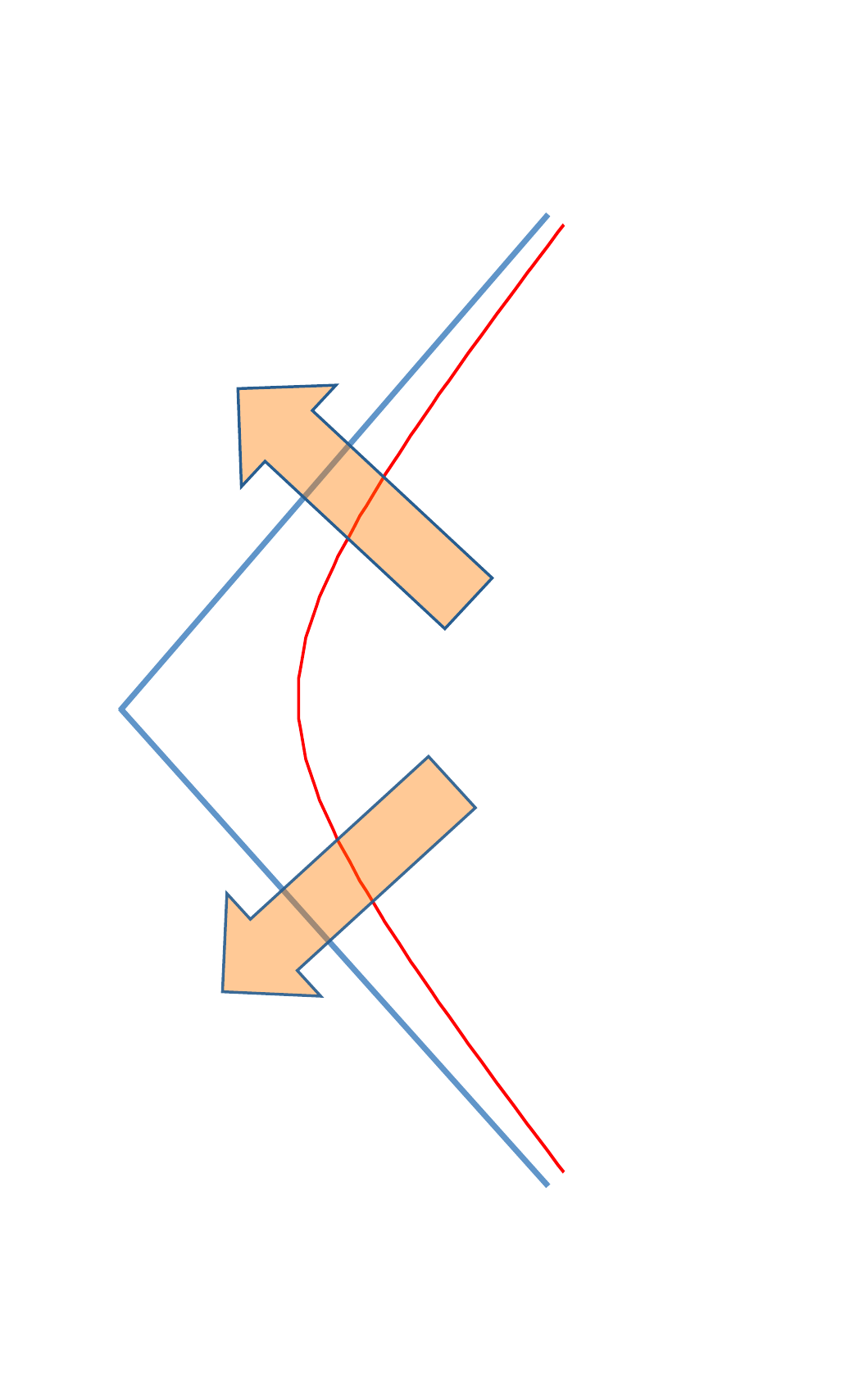}
\caption{Formal direction of the heat flux $\;\dbar Q$.}
\end{center}
\end{figure}

Now compute (note the two minus signs cancel):
\begin{equation}
\fl
{\dbar Q\over\d t} =  \int  \left\{ [T_{00}+T_{33}] \sinh a \tau \cosh a\tau + T_{03} [\sinh^2a\tau +\cosh^2a\tau]  \right\}  {\d\tau\over\d t} \; \d^2 \A.
\end{equation}
Substituting
\begin{equation}
\sinh(a\tau) = a t; \qquad \cosh(a\tau) = \sqrt{1+(at)^2},
\end{equation}
and
\begin{equation}
\cosh(a\tau)\d\tau = \d t; \qquad 
{\d\tau\over\d t}= {1\over\sqrt{1+(a t)^2}},
\end{equation}
we see that we have
\begin{equation}
\fl
{\dbar Q\over\d t} =  \int \left\{ [T_{00}+T_{33}]   a t \sqrt{1+(a t)^2 }+ 
2 T_{03} [2+(a t)^2]  \right\}  {1\over\sqrt{1+(at)^2}}  \d^2 \A,
\end{equation}
thereby implying
\begin{equation}
{\dbar Q\over\d t} =  a \int
\left\{ [T_{00}+T_{33}]  t + 2 T_{03} \left[ {2+(a t)^2\over a\sqrt{1 +(a t)^2 } }\right] \right\}  \d^2\A.
\end{equation}

If at this stage we let $a$ become large (this is mathematically somewhat ill-advised, but close to Jacobson's original construction) then  
\begin{eqnarray}
{\dbar Q/\d t} &\to&   a \;\int   \left\{ [T_{00}+T_{33}] \; t + 2 T_{03} \; |t| \right\} \d^2\A  + \O(1/a)
\nonumber\\
 &=& a \;t\;\int   \left\{ [T_{00}+T_{33}]  + 2 T_{03} \; \sign(t) \right\} \d^2\A + \O(1/a).
\end{eqnarray}
Note that the 2-d integral is to be evaluated on the transverse 2-plane (the $x$-$y$ plane) at time $t$. 
Now defining the null vectors
\begin{equation}
k_\pm^a = \left(1; 0,0, \sign(t) \right),
\end{equation}
which are the null normals on the two segments of the null surface, 
we have
\begin{equation}
\dbar Q/\d t =   a\; t\;  \int \left\{ T_{ab} \; k_\pm^a k_\pm^b\right\} \; \d^2\A + O(1/a). 
\end{equation}
This is as close as we can get to a direct analogue of Jacobson's equation (2) as presented in reference~\cite{Jacobson}. (Note that because $a$ is merely large, not infinite, we are still dealing with timelike trajectories and timelike observers.)

It is mathematically safer to instead proceed in a slightly different manner as follows:
Invoking the Unruh effect, relating the temperature $T$ to magnitude of the 4-acceleration $a$, and explicitly using $k_B T = \hbar a/(2\pi)$, we have
\begin{equation}
\fl
{\dbar Q /\d t\over T} = {2\pi k_B\over\hbar}\; {\dbar Q/\d t \over a} =  {2\pi k_B\over\hbar}\; \int \left\{ [T_{00}+T_{33}] t + 2 T_{03} \left[ {2+(a t)^2\over a\sqrt{1 +(at )^2 } }\right]\right\}  \d^2 \A.
\end{equation}
The key point is that this quantity now has a completely well-defined limit as $a\to\infty$. 

Indeed
\begin{eqnarray}
{\dbar Q/\d t \over T} &\to&   {2\pi k_B\over\hbar}\;\int   \left\{ [T_{00}+T_{33}] \; t + 2 T_{03} \; |t| \right\} \d^2\A 
\nonumber\\
 &=& {2\pi k_B\over\hbar}\;t\;\int   \left\{ [T_{00}+T_{33}]  + 2 T_{03} \; \sign(t) \right\} \d^2\A.
\end{eqnarray}
Therefore
\begin{equation}
{\dbar Q/\d t \over T} \to   {2\pi k_B\over\hbar}\; t\;  \int \left\{ T_{ab} \; k_\pm^a k_\pm^b\right\} \; \d^2\A. 
\end{equation}
This is a mathematically safer version of Jacobson's equation (2). 
It is important to realise this is an \emph{exact} result, valid \emph{globally} for all time. 
Note that both $\,\dbar Q$ and $T$ are diverging as $a\to \infty$, while the ratio $\d S = \dbar Q/T$ remains finite. That is
\begin{equation}
{\d S\over\d t} =   {2\pi k_B\over\hbar}\; t\;  \int \left\{ T_{ab} \; k_\pm^a k_\pm^b\right\} \; \d^2\A. 
\end{equation}
Under normal circumstances the null energy condition [NEC] is satisfied~\cite{twilight}, then $\left\{ T_{ab} \; k_\pm^a k_\pm^b\right\} \geq 0$ and the flux is inwards and positive for $t>0$. On the other hand, the inward flux is negative for $t<0$, indicating that it should be reinterpreted as a positive outward flux.  
That is: The NEC implies a variant of the GSL (generalized second law) holds for this version of Clausius entropy.
\begin{figure}[!htbp]
\label{F:flux-1}
\begin{center}
\includegraphics[scale=0.60, trim = 50 50 50 50]{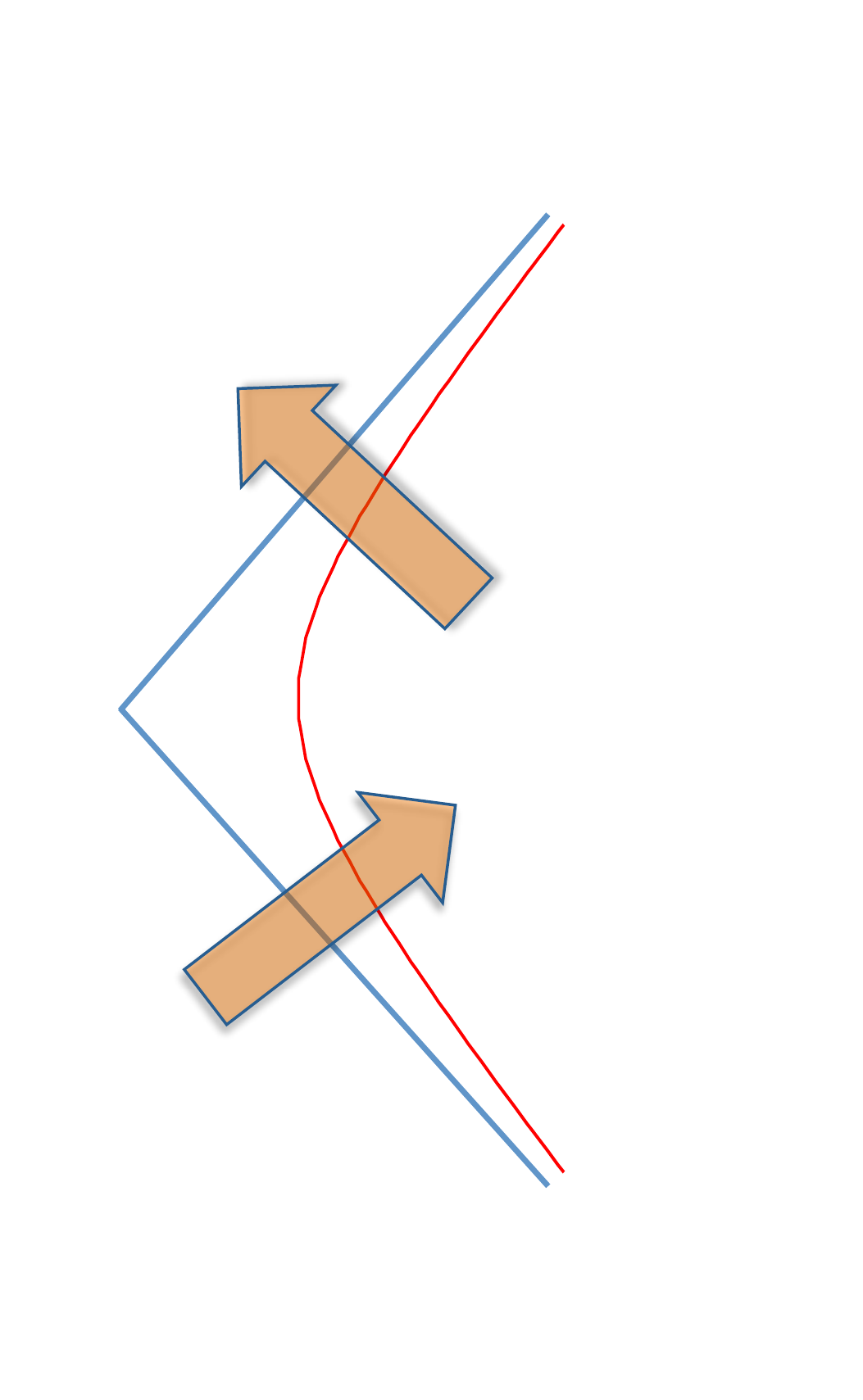}
\caption{Physical direction of the heat flux $\;\dbar Q$, (and the entropy flux $\d S$), assuming the GSL, (which is implied by the NEC),  holds. Assuming the GSL, entropy can only emerge from the past null sheet and enter the future null sheet.}
\end{center}
\end{figure}

The only potentially naively unexpected part of this result is that it is explicitly linear in $t$. Technically that feature can ultimately be traced back to three facts: 
\begin{enumerate}
\item That the location of the bifurcation 2-surface picks out a particular preferred origin for the time coordinate.
\item That symmetry enforces the flux to be zero at the bifurcation 2-surface.
\item That:
\begin{equation}
{1\over a} \; {\d \tau \over \d t} = {\sqrt{1+(at)^2}\over a}  = \sqrt{ {1\over a^2} + t^2} \to |t|.
\end{equation}
\end{enumerate}
Note that the limiting procedure is utterly essential to get the explicit factor of $t$ above. Also, the use of the limiting procedure (starting from a timelike sheet)  is needed for us to be able to invoke the Unruh effect --- since the Unruh effect really makes sense only for timelike observers.
We can now unambiguously write down Clausius entropy \emph{differences} for arbitrary times (both positive or both negative) on the Rindler sheets:
\begin{equation}
\Delta S_\mathrm{Clausius}(t_1,t_2) =  {2\pi k_B\over\hbar}\; \int_{t_1}^{t_2} \bar t\;  \int T_{ab}(\bar t,x,y) \; k_\pm^a k_\pm^b\; \d^2\A \; \d\bar t.
\end{equation}
Perhaps more tellingly we can (in Minkowski space) usefully define the Clausius entropy of the Rindler wedge at time $t$ as:
\begin{equation}
S_\mathrm{Clausius}(t) =  S_\B + {2\pi k_B\over\hbar}\; \int_0^t \bar t\;  \int T_{ab}(\bar t,x,y) \; k_\pm^a k_\pm^b\; \d^2\A \;\d\bar t,
\end{equation}
where $S_\B$ is the entropy to be associated with the bifurcation 2-plane itself, a quantity which is \emph{not} constrained by the current argument.
Again we emphasise that this variant of Jacobson's equation (2) is an \emph{exact} result, valid \emph{globally} for all time. 

We shall now bootstrap this construction away from exact Rindler horizons in flat Minkowski space. We shall first deal with more complicated causal null surfaces in Minkowski space, and then extend the discussion to curved spacetimes.  

\subsection{Causal null cones}\label{SS:cone}

It is now easy to see that the construction above is not limited to Rindler wedges and flat null sheets. (Which is why we spent so much time on the explicit calculation above.) Consider for instance causal null cones defined as follows: 
Choose a spacelike 2-sphere of radius $r_0$, with attached light cones expanding to both future and past.  
Adopt spherical polar coordinates so that the spacelike 2-sphere is
\begin{equation}
x^a(\theta,\phi) =  (0; \;r_0 ,\theta,\phi),
\end{equation}
while the null surface is:
\begin{equation}
x^a(t;\theta,\phi) =  (t; \;r_0 + |t|,\theta,\phi).
\end{equation}
Now pick a spherical sheet of timelike observers
\begin{equation}
x^a(\tau,\theta,\phi) =  \left( {1\over a}\sinh(a\tau); \; r_0+ {1\over a}  \cosh(a\tau), \theta,\phi \right),
\end{equation}
now with 4-velocity
\begin{equation}
V^a(\tau,\theta,\phi) = \left( \cosh(a\tau); \;   \sinh(a\tau), 0,0 \right); \qquad ||V|| = 1,
\end{equation}
and 4-acceleration
\begin{equation}
A^a(\tau,\theta,\phi)=  a \left( \sinh(a\tau); \; \cosh(a\tau),0,0 \right); \qquad ||A|| = a,
\end{equation}
and 4-normal
\begin{equation}
n^a(\tau,\theta,\phi) = -\left( \sinh(a\tau); \;  \cosh(a\tau),0,0 \right); \qquad ||n|| = 1.
\end{equation}
Then
\begin{equation}
\dbar Q =  \; -\int \left(  r_0+ {1\over a}  \cosh(a\tau) \right)^2 T_{ab} \; V^a n^b  \;\d\tau \;\d^2\Omega.
\end{equation}
Note that this is an \emph{inwards} entropy flux;  towards the null cone.
\begin{figure}[!htbp]
\label{F:null-cone}
\begin{center}
\includegraphics[width=10cm, height=6cm,  trim = 50 50 50 50]{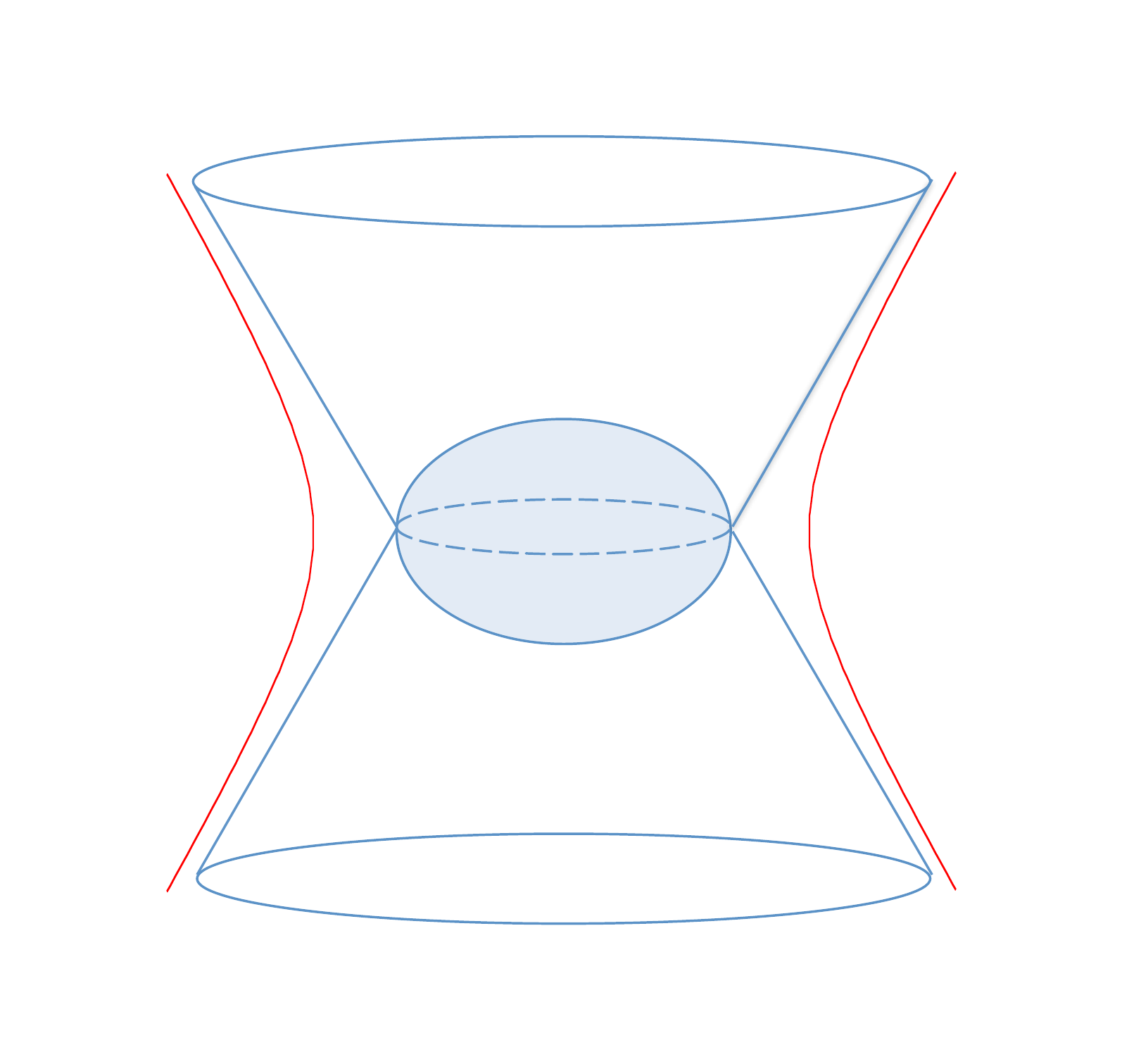}
\caption{Bifurcate double null cone based on a spherical bifurcation 2-surface.\\
Typical timelike observers indicated by red lines.}
\end{center}
\end{figure}

\noindent
Now compute:
\begin{eqnarray}
\fl
{\dbar Q\over\d t}  &=&  \int \left(  r_0+ {1\over a}  \cosh(a\tau) \right)^2 
\nonumber\\
\fl
&&\times 
\left\{ [T_{00}+T_{11}] \sinh a \tau \cosh a\tau + T_{01} [\sinh^2a\tau +\cosh^2a\tau]  \right\}  {\d\tau\over \d t} \;\d^2\Omega.
\end{eqnarray}
Strategically it is now best to substitute
\begin{equation}
\sinh(a\tau) = a t; \qquad \cosh(a\tau) =\sqrt{1+(a t)^2}.
\end{equation}
and
\begin{equation}
\cosh(a\tau) \d\tau = \d t; \qquad \d\tau= {\d t\over\sqrt{1+(a t)^2}}.
\end{equation}
Then
\begin{eqnarray}
\fl
{\dbar Q\over \d t} &=&  \int  \left( r_0+ { \sqrt{1+(a t)^2} \over  a} \right)^2 
\nonumber\\
\fl
&&\times
\left\{ [T_{00}+T_{11}] a t\sqrt{1+(at)^2} + T_{01} [1+2(at)^2]  \right\}  {1\over\sqrt{1+(at)^2}} \; \d^2\Omega,
\end{eqnarray}
and so
\begin{equation}
\fl
{\dbar Q\over \d t} =  \int   \left( r_0+ { \sqrt{1+(at)^2} \over a} \right)^2 
\left\{ [T_{00}+T_{11}] a t  + T_{01} {[1+2(at)^2]\over  \sqrt{1+(at)^2}}
  \right\}  \d^2\Omega.
  \end{equation}
Therefore, again invoking the Unruh effect,
\begin{eqnarray}
\fl
{\dbar Q/\d t \over T} &=&  {2\pi k_B\over\hbar}\;  {\dbar Q/\d t \over a} 
\\
\fl
&=& 
 {2\pi k_B\over\hbar}\;  \int 
\left( r_0+ { \sqrt{1+(at)^2} \over  a} \right)^2 
\left\{ [T_{00}+T_{11}]  t  + T_{01} {[1+2(at)^2]\over  a \sqrt{1+(at)^2}}
  \right\} \d^2\Omega.
\end{eqnarray}
This quantity now has a well-defined limit as $a\to\infty$. 
Indeed we have the \emph{exact} result
\begin{eqnarray}
{\dbar Q/\d t \over T} &\to&    {2\pi k_B\over\hbar}\;
\int   \left( r_0+  |t|  \right)^2 \left\{ [T_{00}+T_{11}] \; t  + 2 T_{01}\; |t| \right\} \d^2\Omega
\\
 &=&   {2\pi k_B\over\hbar}\;   \left( r_0+ |t| \right)^2  t \; \int \left\{ T_{ab} \; k_\pm^a k_\pm^b  \right\}  \d^2\Omega.
\end{eqnarray}
Here
\begin{equation}
k_\pm^a = (1; \sign(t), 0,0),
\end{equation}
and the angular integral is to be carried out over the 2-sphere at time $t$. Pulling the factor $ \left( r_0+ |t| \right)^2$ inside the integral we obtain the \emph{exact} result
\begin{equation}
{\dbar Q/\d t \over T} \to  {\d S\over\d t} =  {2\pi k_B\over\hbar}\;    t \; \int \left\{ T_{ab} \; k_\pm^a k_\pm^b  \right\}  \d^2\A.
\end{equation}
The integral is now over the area of the 2-sphere at time $t$. Formally the final result is completely equivalent to that obtained for the Rindler wedge, even though various intermediate steps were somewhat different. This observation is particularly important, in that it will now allow us to greatly extend the range of validity of our previous result. In particular for any causal null cone (light cone) we now have
\begin{equation}
S_\mathrm{Clausius}(t) =  S_\B + 
{2\pi k_B\over\hbar}\; \int_0^t \bar t\;  \int T_{ab}\left(\bar t,\x(\theta,\phi,\bar t\,)\right) \; k_\pm^a k_\pm^b\; \d^2\A \;\d\bar t,
\end{equation}
Here $S_\B$ is now the Clausius entropy to be associated with the bifurcation 2-sphere of radius $r_0$ located at $t=0$. 
\begin{figure}[!htbp]
\label{F:null-cone}
\begin{center}
\includegraphics[width=10cm, height=6cm,  trim = 50 50 50 50]{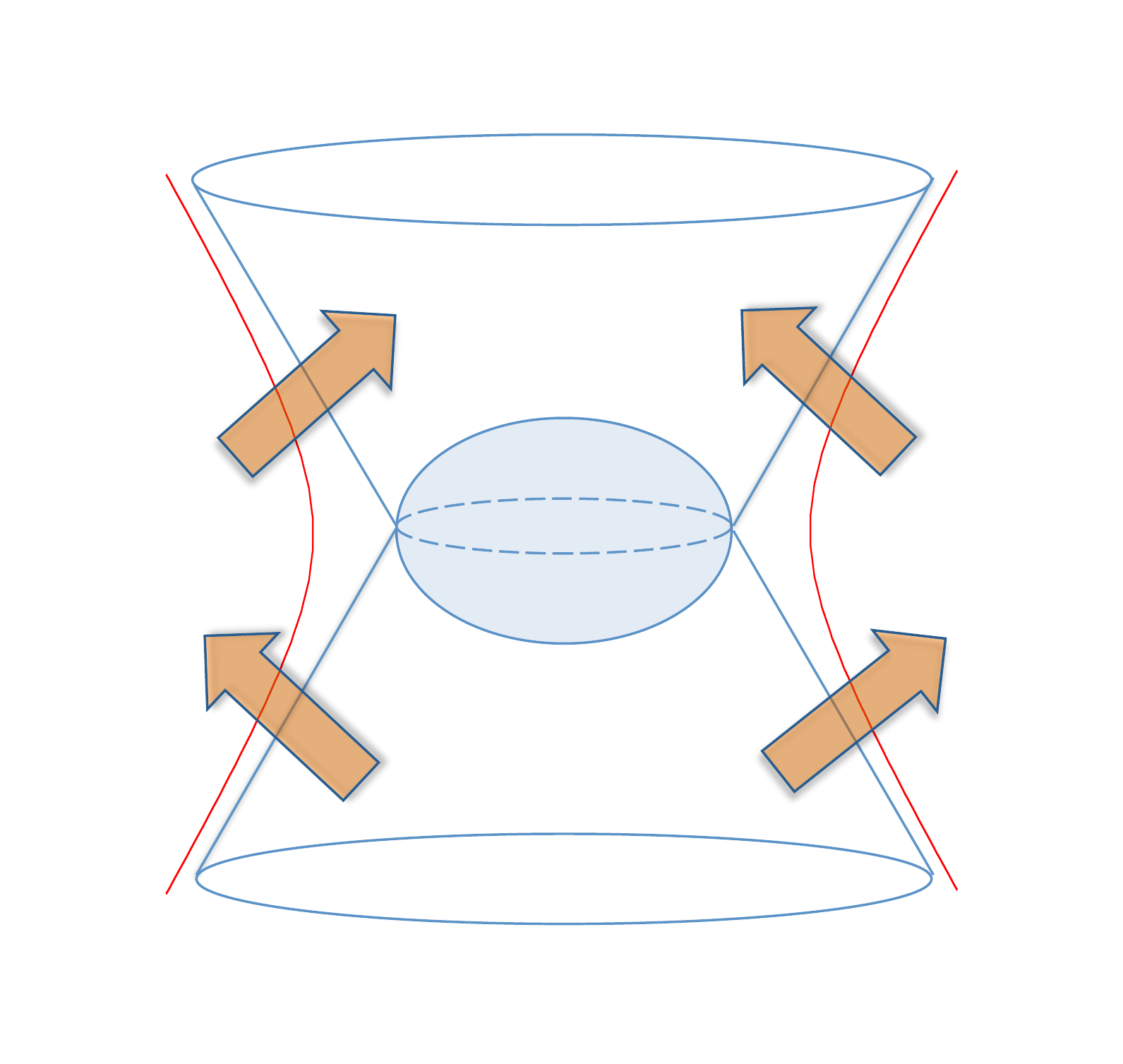}
\caption{Direction of physical entropy fluxes, (assuming the GSL, which is implied by the NEC), for the bifurcate double null cone based on a spherical bifurcation 2-surface.  Assuming the GSL, entropy can only emerge from the past null cone and enter the future null cone.}
\end{center}
\end{figure}

\subsection{Convex-base null conoids}\label{SS:conoid}

Consider now an arbitrary convex spacelike 2-surface. Choose Cartesian coordinates $x^a$ in Minkowski space, and generic coordinates $\xi^i$ on the 2-surface. Then we can write
\begin{equation}
x^a(\xi^i) =  \left(0; \;\x(\xi^i)\right).
\end{equation}
Because the surface is convex, its outward pointing normals $\n(\xi)$ never intersect, so we can attach outward pointing past and future light rays to each point on the surface, and in turn these light rays will never intersect --- so they define null surfaces. Then for these null sheets
\begin{equation}
x^a(t;\xi^i) =  \left(t; \;\x(\xi^i)+|t|\;\n(\xi^i) \right).
\end{equation}
The resulting null conoids intersect at the original spacelike 2-surface, which is therefore a bifurcation 2-surface. Along each one of these normal directions we can now simply copy the calculation for causal null cones as presented above --- which is why we put the effort into an explicit calculation for those simple cases.  We again see
\begin{equation}
{\dbar Q/\d t \over T} \to  {\d S\over\d t} = {2\pi k_B\over\hbar}\;    t \; \int \left\{ T_{ab} \; k_\pm^a k_\pm^b  \right\}  \d^2\A.
\end{equation}
The integral is now over the cross-sectional area of the conoid at time $t$. The result is again \emph{exact} and valid \emph{globally} for all time. 
We now see
\begin{equation}
S_\mathrm{Clausius}(t) =  S_\B + {2\pi k_B\over\hbar}\; \int_0^t \bar t\;  \int T_{ab}(\bar t,\x(\xi,t)) \; k_\pm^a k_\pm^b\; \d^2\A \;\d\bar t,
\end{equation}
with the integral running over the null conoid.

\subsection{Causal diamonds}\label{SS:diamond}

To understand what happens if the bifurcation 2-surface is concave, (even partially concave), it is best to start with the highly-symmetric causal diamond configuration. 
\begin{figure}[!htbp]
\label{F:causal-diamond}
\begin{center}
\includegraphics[width=4cm, height=6cm,   trim = 50 50 50 50]{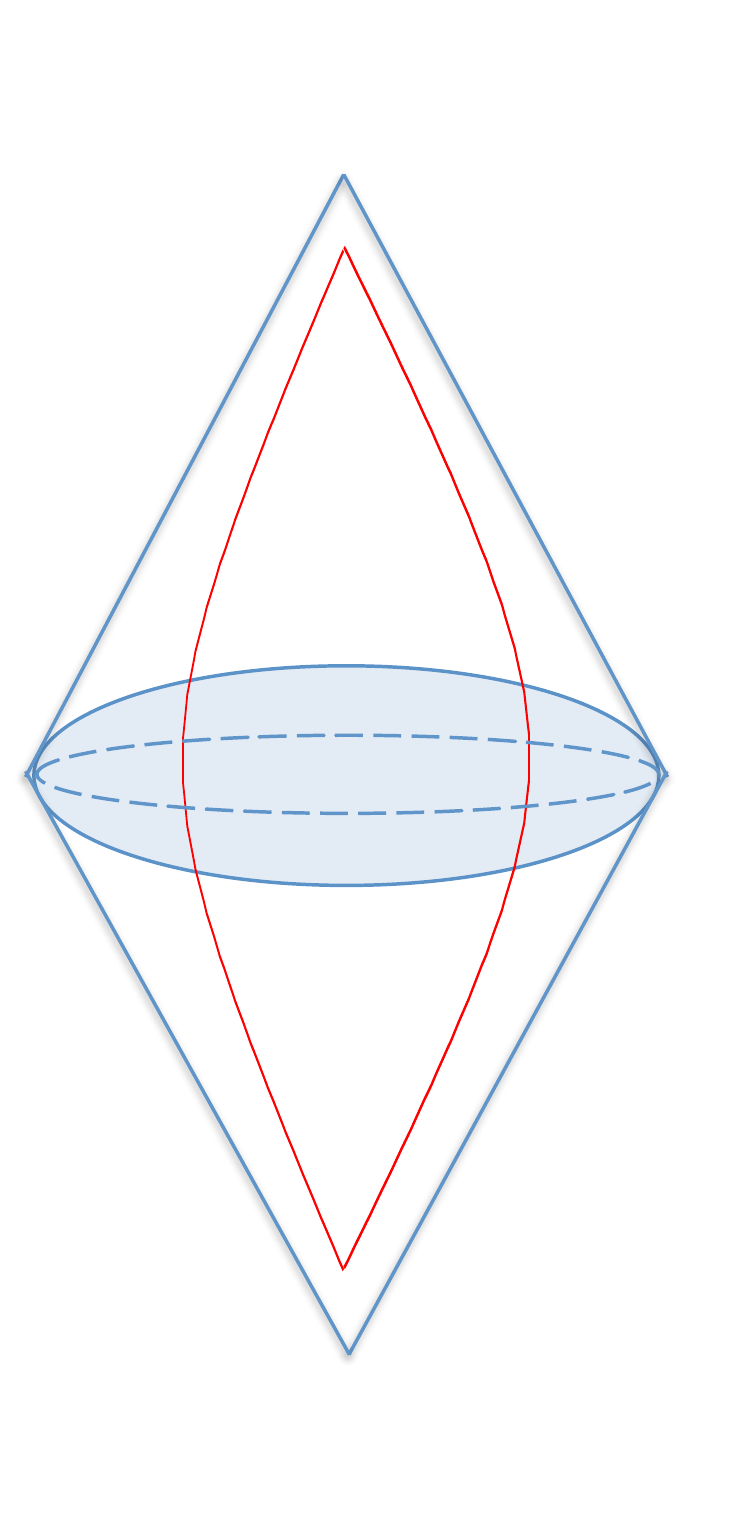}
\caption{Causal diamond configuration.\\
Typical timelike observers indicated by red lines.\\
Note timelike observers now collide at finite time.}
\end{center}
\end{figure}

\noindent
Choose a spacelike 2-sphere of radius $r_0$, but now with attached light cones \emph{contracting} to both future and past.  
The null surface is now
\begin{equation}
x^a(t;\theta,\phi) =  (t; \;r_0 - |t|,\theta,\phi).
\end{equation}
The null curves generating the null surface now all collide at two points at $t_\mathrm{collision}=\pm r_0$. A suitable class of timelike observers is now
\begin{equation}
x^a(\tau,\theta,\phi) =  \left( {1\over a}\sinh(a\tau); \; r_0- {1\over a}  \cosh(a\tau), \theta,\phi \right),
\end{equation}
with the timelike observers colliding at 
\begin{equation}
\tau_\mathrm{collision}  = \pm{1\over a} \; \cosh^{-1}(r_0 a),
\end{equation}
corresponding to 
\begin{equation}
t_\mathrm{collision}=\pm \sqrt{r_0^2-{1\over a^2}}.
\end{equation}
As long as we restrict attention to the finite interval where these timelike curves do not intersect then the previous computation for causal null cones can be carried over, and taking the appropriate limit we see that in the finite interval $t\in(-r_0,r_0)$ we still have
\begin{equation}
{\dbar Q/\d t \over T} \to  {\d S\over\d t} =  {2\pi k_B\over\hbar}\;    t \; \int \left\{ T_{ab} \; k_\pm^a k_\pm^b  \right\}  \d^2\A.
\end{equation}
The only minor quirk is that the timelike observers now reside \emph{inside} the null surface, and that the timelike observers have 4-normal
\begin{equation}
n^a(\tau,\theta,\phi) = \left( -\sinh(a\tau); \;  \cosh(a\tau),0,0 \right); \qquad ||n|| = 1.
\end{equation}
This implies one is now calculating an \emph{outward} entropy flux. If one assumes the NEC this corresponds to a positive outwards flux for $t>0$ and a negative outwards (positive inwards) flux for $t<0$, which is compatible with the GSL.  Again
\begin{equation}
S_\mathrm{Clausius}(t) =  S_\B + 
{2\pi k_B\over\hbar}\; \int_0^t \bar t\;  \int T_{ab}\left(\bar t,\x(\theta,\phi,\bar t\,)\right) \; k_\pm^a k_\pm^b\; \d^2\A \;\d\bar t.
\end{equation}
Though historically Jacobson's construction was first applied to Rindler horizons, (and curved space local Rindler horizons), 
the causal diamond construction, (and its curved space analogue), can plausibly be argued to be more natural. In particular the causal diamond construction makes it clear that ``local'' causal horizons are already sufficiently interesting --- there is no need to continue the causal surfaces of interest all the way to (past or future) null infinity. 

\begin{figure}[!htbp]
\label{F:causal-diamond}
\begin{center}
\includegraphics[width=6cm, height=8cm,   trim = 50 50 50 50]{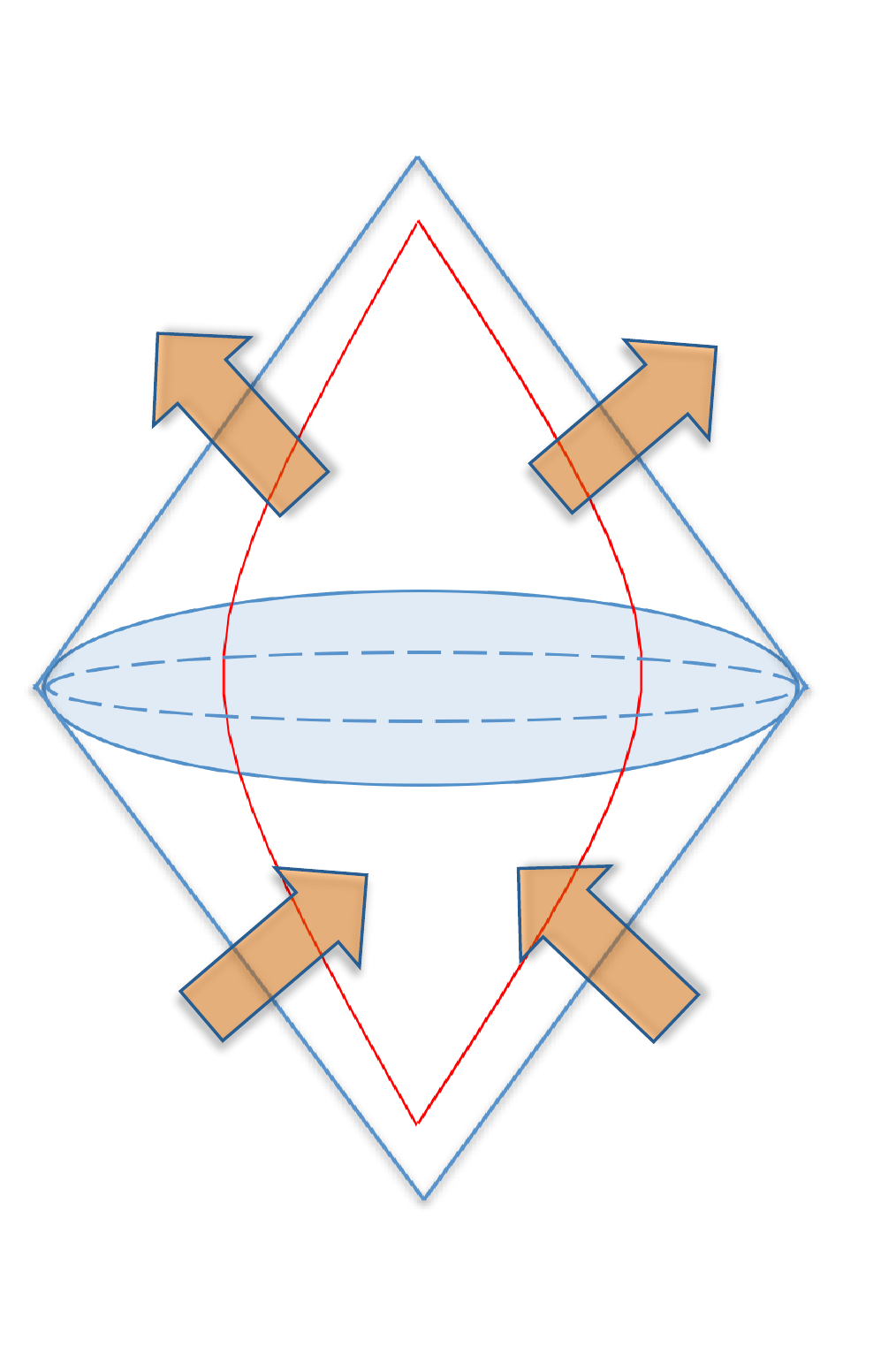}
\caption{Direction of the physical entropy fluxes, (assuming the GSL, which is implied by the NEC), for the causal diamond configuration. Assuming the GSL, entropy can only enter the causal diamond  from the past null cone and leave the causal diamond via  the future null cone.}
\end{center}
\end{figure}

\subsection{Generic null conoids}\label{S:generic-conoids}
Now consider arbitrary null conoids  in flat Minkowski space.  
We start with some arbitrary 2-surface at time $t=0$,
\begin{equation}
x^a(\xi^i) =  \left(0; \;\x(\xi^i)\right),
\end{equation}
but now with no constraint on the convexity of the 2-surface. 
All the real work has already been done --- the only obstruction comes from intersecting null normals. 
We see that over some finite interval $t\in(-t_*,t_*)$, where $t_*$ is determined by the time of first intersection of the null normals, we still have the \emph{exact} result
\begin{equation}
{\dbar Q/\d t \over T} \to  {\d S\over\d t} =  {2\pi k_B\over\hbar}\;    t \; \int \left\{ T_{ab} \; k_\pm^a k_\pm^b  \right\}  \d^2\A.
\end{equation}
Consequently, for the Clausius entropy, at as long as $t\in(-t_*,t_*)$,  we still have the \emph{exact} result
\begin{equation}
S_\mathrm{Clausius}(t) =  S_\B 
+ {2\pi k_B\over\hbar}\; \int_0^t \bar t\;  \int T_{ab}\left(\bar t,\x(\xi,\bar t\,)\right) \; k_\pm^a k_\pm^b\; \d^2\A \;\d\bar t.
\end{equation}
(Note again that we do not need to extrapolate the null causal surfaces all the way to past or future null infinity in order to have an interesting notion of Clausius entropy. Indeed, in the presence of null caustics, such an extension might, apart from being unnecessary, be outright impossible.)
While the discussion started out with a straightforward computation for Rindler horizons, we have now bootstrapped it to a large class of bifurcate null surfaces (still in Minkowski space) --- the only limitation at this stage is that there be an inertial frame in which the bifurcation surface can be chosen to lie on the hyperplane $t=0$.

\subsection{Generic bifurcate null surfaces}\label{S:bifurcate}

As a penultimate step, we are now ready to address the situation for generic bifurcate null sheets in flat Minkowski space.   We start with some completely arbitrary spacelike 2-surface,
\begin{equation}
x^a(\xi^i) =  \left(t_0(\xi^i); \;\x(\xi^i)\right),
\end{equation}
but now with no constraint on the convexity of the 2-surface, nor with any constraint that the 2-surface be contained in a hyperplane. 
Picking normals $\n(\xi)$ to the spatial part of this 2-surface, so we can attach outward pointing past and future light rays to each point on the surface --- thereby  defining null surfaces. Then for these null sheets one convenient parameterization is
\begin{equation}
x^a(t;\xi^i) =  \left( t_0(\xi^i) + t; \;\x(\xi^i)+|t|\;\n(\xi^i) \right) = x^a(\xi^i) + t \; k_\pm^a(\xi^i).
\end{equation}
Again, all the real work has already been done --- the only significant obstruction comes from intersecting null normals. 
We see that over some finite interval $t\in(-t_*,t_*)$, where $t_*$ is determined by the time of first intersection of the null normals, we still have the \emph{exact} result
\begin{equation}
{\dbar Q/\d t \over T} \to   {\d S\over\d t} = {2\pi k_B\over\hbar}\;    t \; \int \left\{ T_{ab}\left( x(\xi) + t \; k_\pm(\xi) \right) \;  k_\pm^a k_\pm^b  \right\}  \d^2\A.
\end{equation}
One subtlety is that this is not precisely $\d S/\d t$ ``at physical time $t$'';  rather this  $\d S/\d t$ is obtained by propagating the bifurcation surface $\B$ forward by time $t$ in some arbitrarily chosen rest frame and calculating the flux as a function of this evolution parameter.
Consequently, for the Clausius entropy we can still write down an \emph{exact} result
\begin{equation}
S_\mathrm{Clausius}(t) =  S_\B 
+ {2\pi k_B\over\hbar}\; \int_0^t \bar t\;  \int T_{ab}\left(x(\xi) +\bar t \; k_\pm(\xi) \right) \; k_\pm^a k_\pm^b\; \d^2\A \;\d\bar t.
\end{equation}
As a final step we note that we could independently pick distinct affine parameters $\lambda$ on each null generator and write
\begin{equation}
x^a(\lambda;\xi^i) =  x^a(\xi^i) + \lambda \; k_\pm^a(\xi^i); \qquad k^a_\pm(\xi^i) = {\d x^a(\lambda,\xi^i)\over\d \lambda}.
\end{equation}
Then taking $\S$ to be any spacelike cross-section of the bifurcate null surface we have
\begin{equation}
S_\mathrm{Clausius}(\S) =  S_\B 
+ {2\pi k_B\over\hbar}\; \int_\B^\S \lambda\; T_{ab}\left(x(\xi) +\lambda\; k_\pm(\xi) \right) \; k_\pm^a k_\pm^b\; \d^2\A \;\d\lambda.
\end{equation}
Here the integral now runs over the entire null surface between $\B$ and $\S$.  Note that the construction is manifestly independent of the way the affine parameter is normalized on each null generator.

We emphasize that while the discussion started out with a straightforward computation for exact Rindler horizons, we have now bootstrapped it to essentially arbitrary bifurcate null surfaces (still in Minkowski space).  We shall now perform a consistency check on the reasonableness of the construction, and then generalize the construction to curved spacetimes.

\section{Compatibility with the Bekenstein bound}\label{S:Bekenstein-bound}

Let us now check our proposal for the Clausius entropy for compatibility with the Bekenstein bound~\cite{B-bound}:
\begin{equation}
S \leq  k_B {2\pi M R\over \hbar}.
\end{equation}
This inequality was argued by Bekenstein to apply to weakly bound and weakly interacting systems.  Since our Clausius notion of entropy is at this stage purely a Minkowski space result, the system is certainly weakly bound. But how are we to take this quantity, 
\begin{equation}
S_\mathrm{Clausius}(t) =  S_\B + {2\pi k_B\over\hbar}\; \int_0^t \bar t\;  \int T_{ab}(\bar t,\x(\bar t)) \; k_\pm^a k_\pm^b\; \d^2\A \;\d\bar t,
\end{equation}
and relate it to Bekenstein's bound? Certainly some extra assumptions will be required. (Such as, which bifurcate null surfaces will we consider?)  

Let us first choose the bifurcation surface to be a single point, and the null surface to be its future light cone. 
When the bifurcation surface is a single point it is plausible to set $S_\B\to0$. 
For simplicity, let us first take  the stress-energy to be that of a spherically symmetric perfect fluid, then
\begin{equation}
S_\mathrm{Clausius}(t) =   {2\pi k_B\over\hbar}\; 4\pi \int_0^t \bar t^3\;  (\rho+p) \;\d\bar t.
\end{equation}
Further note that $t\to R$, the radius of the light-sphere at time $t$. (We have set $c\to1$.) 
For the specific case of a constant density fluid we then have
\begin{equation}
S_\mathrm{Clausius}(R) =   {2\pi k_B\over\hbar}\; 4\pi(\rho+p)  {R^4\over 4}.
\end{equation}
But the Bekenstein bound is asserted to apply  to weakly interacting systems, so it is for current purposes acceptable to take $p\in(0,\rho/3)$. This is equivalent to the so-called ``trace energy condition''. The TEC is one of the oldest of the classical energy conditions, which was subsequently abandoned  as fundamental physics, though it is certainly a useful characterization for weakly interacting matter~\cite{twilight}. Under these conditions $\rho+p < {4\over3} \rho$, and so we have
\begin{equation}
S_\mathrm{Clausius}(R) <   {2\pi k_B\over\hbar}\;  {4\pi \rho R^3\over 3}  R 
= k_B {2\pi M  R\over \hbar},
\end{equation}
as required. Consequently the notion of Clausius entropy defined in this article is indeed compatible with the Bekenstein bound. This gives us additional confidence that the construction developed above is physically interesting. 

If the density and pressure are not constant (but are at least spherically symmetric) a minor variant of the above argument considers the quantity
\begin{equation}
\fl
X = 4\pi \int_0^R r^3 (\rho+p) \;\d r < 4\pi \int_0^R  r^3(4\rho/3) \;\d r 
= \int_0^R r \; \d m(r) + {1\over3} \int_0^R r \; \d m(r).
\end{equation}
But then by integration by parts
\begin{equation}
\fl 
X <  M R -  \int_0^R \left[ m(r) \; \d r - {1\over3} r \; \d m(r) \right] = M \; R +{1\over3} \int_0^R r^4 \; {\d[m(r)/r^3]\over\d r}\;  \d r.
\end{equation}
If we now assume the average density is decreasing as one moves outwards, then $\d[ m(r)/r^3]/\d r < 0$, and the last term is negative. This falloff condition on the average density is one of the specific conditions Chandrasekhar uses in his investigations of non-relativistic stellar structure~\cite{stellar}. Then $X< MR$, and we again see that our construction for the Clausius entropy is at least compatible with Bekenstein's bound for weakly interacting systems.  

\section{Curved spacetime}\label{S:curved}

Now that we have carried out this exact calculation for flat Minkowski space, and checked for compatibility with wider notions of what we expect entropy to be, the generalization to curved spacetime is straightforward.

\subsection{Near the bifurcation 2-surface}\label{SS:near}

First, consider an approximate calculation for curved spacetime in the vicinity of the bifurcation 2-surface. Pick a bifurcate null surface in some curved spacetime. Pick a point  on that bifurcation 2-surface. 
In the vicinity of that point adopt Gaussian normal coordinates $x^a$ so that 
\begin{equation}
g_{ab} = \eta_{ab} + \O([x^a]^2).
\end{equation}
Then to the appropriate level of accuracy the null curves emanating from this point on the bifurcation 2-surface can be represented by
\begin{equation}
x^a(t) =  \left(t; \; 0,0, |t| \right) +  \O( t^2).
\end{equation}
An appropriate timelike observer is
\begin{equation}
x^a(\tau)  = \left( {1\over a}\sinh(a\tau); \; 0,0,  {1\over a}  \cosh(a\tau) \right) +   \O( \tau^2).
\end{equation}
Equivalently
\begin{equation}
x^a(t)  = \left( t ; \; 0,0, \sqrt{t^2+{1\over a^2}}\right) +   \O( t^2).
\end{equation}
Differentiating, the 4-velocity and 4-normal are determined up to terms of $\O(t)$, and the 4-acceleration up to terms of $\O(1)$. Furthermore, note that $T_{ab}(t) = T_{ab}(0) + \O(t)$. 

Finally, although the timelike observers are no longer exactly hyperbolic for all time, there is an adiabatic argument~\cite{Barbado:2012} demonstrating that the the Unruh effect will still hold adiabatically as long as the region over which the motion is close to hyperbolic, (the size of this region being determined by the spacetime curvature), is large compared to the distance scale $1/a$.  We emphasise that there is now a considerable body of work on what might be called the ``finite-time Unruh effect'', wherein the original simplifying assumptions of eternal-constant-acceleration observers~\cite{Unruh:1976} is dispensed with. See for instance references~\cite{Sriramkumar:1994, Schlicht:2003, Satz:2006, Louko:2006, Obadia:2007, Kothawala:2009, Doukas:2013}. 
(Similarly, in a black hole situation the existence  of the Hawking effect is not dependent on the presence of an exact stationary [event] horizon, an approximate horizon satisfying a suitable adiabaticity condition is quite sufficient for the emission of a Planckian spectrum of Hawking photons~\cite{Visser-nas, Barcelo-nas1, Barcelo-nas2, Barcelo-nas3}.)

Inserting all this into the previous computation, and taking the limit $a\to\infty$, we now get the \emph{approximate} result
\begin{equation}
{\dbar Q/\d t \over T} \to  {\d S\over\d t} =  {2\pi k_B\over\hbar}\;    t \; \int_\B\left\{ T_{ab}(0) \; k_\pm^a k_\pm^b  \right\}  \d^2\A + \O(t^2),
\end{equation}
where to the relevant level of approximation the integral now runs over the bifurcation 2-surface $\B$. 
A subtlety here is that the Gaussian normal coordinate construction implies that one is free to choose the $t$ coordinate independently on each null generator of the bifurcate null surface. This is equivalent to the ability to choose an arbitrary affine parameter $\lambda$ for each null generator, and to make this more explicit we can write
\begin{equation}
{\dbar Q/\d \lambda \over T} \to   {\d S\over \d\lambda} = {2\pi k_B\over\hbar}\;    \lambda \; \int_\B\left\{ T_{ab}(0) \; k_\pm^a k_\pm^b  \right\}  \d^2\A + \O(\lambda^2).
\end{equation}
If we restrict attention to ``locally Rindler'' bifurcate null surfaces then this expression is one of the key steps in Jacobson's thermodynamic derivation of the Einstein equations~\cite{Jacobson} --- this is essentially Jacobson's equation (2) --- but it is now clear from the present discussion that at least this aspect of Jacobson's argument is much more general, applying to essentially arbitrary bifurcate null surfaces.  Note that this is the inward entropy flux. For positive $t$ and matter satisfying the NEC the flux is positive inwards. The sign flip for negative $t$ indicates the entropy flow is then positive outwards. 
Consequently, for the Clausius entropy we now have 
\begin{equation}
S_\mathrm{Clausius}(t) =   S_\B + {2\pi k_B\over\hbar}
\; {t^2\over 2}  \int_\B \left\{ T_{ab}(0) \; k_\pm^a k_\pm^b\right\}  \d^2\A + \O(t^3).
\end{equation}
The constant term $S_\B$ is again undetermined by this argument. 
It is to be emphasized that this construction is to be applied to bifurcate null surfaces, not spacelike volumes, and the construction is both qualitatively and quantitatively different from entropy estimates built up by integrating up the thermodynamic entropy associated with small individual lumps of matter~\cite{Abreu1, Abreu2, Abreu3, Abreu4}.
If we choose to work with arbitrary affine parameters and arbitrary spacelike sections $\S$ of the bifurcate null surface then we can rewrite the result as 
\begin{equation}
S_\mathrm{Clausius}(\S) =   S_\B + {2\pi k_B\over\hbar}
\;  \int_\B \left\{ {\lambda^2\over 2} \; T_{ab}(0) \; k_\pm^a k_\pm^b\right\}  \d^2\A + \O(\lambda^3).
\end{equation}
Here $\S$ is now the 2-surface defined by propagating an affine distance $\lambda$ along each null generator emanating from the bifurcation 2-surface $\B$. 

\subsection{General formula for curved-space Clausius entropy}\label{SS:general}

In view of the above discussion we can now simply \emph{postulate} that for arbitrary bifurcate null surfaces in curved spacetime, at arbitrary cross-section $\S$ of  the null surface
\begin{equation}
S_\mathrm{Clausius}(\S) \equiv  S_\B 
+ {2\pi k_B\over\hbar}\; \int_\B^\S \lambda \; T_{ab}\left(x(\xi,\lambda) \right) \; k_\pm^a k_\pm^b\; \d^2\A \;\d\lambda.
\end{equation}
Note that $\lambda$ is an affine null parameter, that this integral is well-defined in the sense that it is invariant under rescaling of the affine null parameter, and that in view of the preceding discussion this construction  passes all the consistency tests one might reasonably wish to impose. The only real restriction on the construction is that one should stop using it as soon as the null surface develops self-intersections. 

\subsection{Generalized second law}\label{SS:GSL}
Note in particular that imposing the classical null energy condition --- the NEC --- would guarantee positivity of the Clausius entropy flux, and imply a version of the GSL. Thus the NEC is a \emph{sufficient} condition for the GSL to hold. (While there are certainly quantum-induced violations of the energy conditions~\cite{twilight}, we would argue that they can be neglected in the thermodynamic limit.) 

Note that a rather weaker \emph{sufficient} condition for the GSL to hold, (for this definition of Clausius entropy),  is that on all closed (or at worst edgeless) spacelike 2-surfaces
\begin{equation}
\int_\S  T_{ab}\left(x(\xi) \right) \; k_\pm^a(\xi) k_\pm^b(\xi)\; \d^2\A(\xi) \geq 0.
\end{equation}
A slightly different \emph{sufficient} condition for the GSL to hold \emph{asymptotically}, (at sufficiently late or early times, for this particular definition of Clausius entropy), is that on all future-pointing null half-geodesics we have
\begin{equation}
\int_0^\infty \lambda \; T_{ab}\left(x(\lambda) \right) \; k_+^a(\lambda) k_+^b(\lambda) \;\d\lambda \geq 0,
\end{equation}
and that on all past-pointing null half-geodesics we have
\begin{equation}
\int_{-\infty}^0  \lambda \; T_{ab}\left(x(\lambda) \right) \; k_-^a(\lambda) k_-^b(\lambda) \;\d\lambda \leq 0.
\end{equation}
These conditions are certainly implied by the NEC, but are  very much weaker than the NEC. Thus the GSL is seen to hold under very much weaker conditions than the NEC.

These integral variants of the NEC are also qualitatively very different from the standard averaged null energy condition (ANEC), see for instance~\cite{wormholes, censor, Flanagan:1996, q-anec}, or the Ford--Roman quantum inequalities~\cite{Ford:1990, Ford+Roman, Ford:1996, Ford+Roman+collaborators}, or their variants~\cite{Ford:1999, Fewster:1999a, Fewster:1999b, Fewster:2007}, or even the recent non-linear energy conditions explored in~\cite{Prado1, Prado2}. This strongly suggests these nonstandard integral variants of the NEC are well worth additional scrutiny.

\section{Discussion}\label{S:discussion}

The net result of this calculation, and the construction it inspires, is that one can associate an observer-dependent notion of entropy, very closely related to the Clausius entropy (thermodynamic entropy,  $\;\dbar Q/T$ entropy)~\cite{Clausius}, and a generalization of Jacobson's local-Rindler entropy~\cite{Jacobson}, to any arbitrary bifurcate null surface. That is, there is a certain sense in which even Clausius entropy ($\,\dbar Q/T$ entropy) is observer-dependent, with a ``virtual Clausius entropy'' being associated with arbitrary bifurcate ``virtual causal horizons''.  (See also, for instance,  the discussion in references~\cite{paddy1, paddy2, paddy3}.)  This construction, because it generalizes one part of Jacobson's ``thermodynamic'' derivation of the Einstein equations, cuts to the heart of the issue of the putative universal equality of thermodynamic and entanglement entropy. We will address such issues more fully in future work. 

\ack

VB acknowledges support by a Victoria University PhD scholarship.
MV acknowledges support via the Marsden Fund, and via a James Cook Fellowship, both grants being administered by the Royal Society of New Zealand.  The authors wish to thank the referees for constructive criticisms and suggestions. 

\section*{References}


\begin{thebibliography}{69}

\bibitem{Jacobson}
Ted Jacobson,
``Thermodynamics of space-time: The Einstein equation of state'',\\
 Phys.\ Rev.\ Lett.\ {\bf 75} (1995) 1260
[gr-qc/9504004].


\bibitem{Clausius}
Clausius, Rudolf (1862). \\
Communicated to the Naturforschende Gesellschaft of Zurich, January 27, 1862; \\
published in the Vierteljahrschrift of this Society, vol. vii. p. 48; \\
in Poggendorff's Annalen, May 1862, vol. cxvi. p. 73; \\
in the Philosophical Magazine, S. 4. vol. xxiv. pp. 81, 201; \\
and in the Journal des Mathematiques of Paris, S. 2. vol. vii. p. 209.

\bibitem{Bekenstein}
Jacob  Bekenstein, ``Black holes and entropy'', Phys. Rev. D 7 (1973) 2333--2346. 

\bibitem{Shannon}
Claude E Shannon, ``A Mathematical Theory of Communication'', \\
Bell System Technical Journal {\bf27} (3) (July/October 1948)  379--423. \\
{\sf http://cm.bell-labs.com/cm/ms/what/shannonday/shannon1948.pdf}

\hspace{-13.35pt}
Claude E Shannon and Warren Weaver, ``The Mathematical Theory of Communication'', \\
University of Illinois Press, 1949. ISBN 0-252-72548-4

\bibitem{vonNeumann}
John von Neumann,  \emph{Mathematische Grundlagen der Quantenmechanik}. \\
(Springer, Berlin, 1955). ISBN 3-540-59207-5; 

\hspace{-13.35pt}
John von Neumann,  \emph{Mathematical Foundations of Quantum Mechanics}. \\
(Princeton University Press, Princeton, 1996). ISBN 978-0-691-02893-4.

\bibitem{Jaynes:1957a}
E T Jaynes, ``Information Theory and Statistical Mechanics'', \\
Physical Review Series II {\bf106} (1957) 620--630. doi:10.1103/PhysRev.106.620.
 
 \bibitem{Jaynes:1957b}
E T Jaynes, ``Information Theory and Statistical Mechanics II'', \\
Physical Review Series II {\bf108} (1957) 171--190. doi:10.1103/PhysRev.108.171.


\bibitem{Bombelli:1986}
  L.~Bombelli, R.~K.~Koul, J.~Lee and R.~D.~Sorkin,\\
  ``A Quantum Source of Entropy for Black Holes'',
  Phys.\ Rev.\ D {\bf 34} (1986) 373.
  
\bibitem{Srednicki:1993}
  M.~Srednicki,
  ``Entropy and area'',
  Phys.\ Rev.\ Lett.\  {\bf 71} (1993) 666
  [hep-th/9303048].
  
  \bibitem{Coleman:1991}
  S.~R.~Coleman, J.~Preskill and F.~Wilczek,
  ``Quantum hair on black holes'',\\
  Nucl.\ Phys.\ B {\bf 378} (1992) 175
  [hep-th/9201059].

  \bibitem{Preskill:1992}
  J.~Preskill,
  ``Do black holes destroy information?'',\\
  In: Houston 1992, Proceedings, Black holes, membranes, wormholes and superstrings, 22--39,
  [hep-th/9209058].
 
\bibitem{Fiola:1994}
  T.~M.~Fiola, J.~Preskill, A.~Strominger and S.~P.~Trivedi,\\
  ``Black hole thermodynamics and information loss in two-dimensions'',\\
  Phys.\ Rev.\ D {\bf 50} (1994) 3987
  [hep-th/9403137].
  

\bibitem{Langer:1990}
  Stephen A Langer, James P Sethna, and Eric R Grannan,\\
  ``Nonequilibrium entropy and entropy distributions'',
  Physical Review B {\bf41} (1990) 2261--2278. 
  
  
  \bibitem{Chirco:2009}
  G.~Chirco and S.~Liberati,\\
  ``Non-equilibrium Thermodynamics of Spacetime: The Role of Gravitational Dissipation'',\\
  Phys.\ Rev.\ D {\bf 81} (2010) 024016
  [arXiv:0909.4194 [gr-qc]].


\bibitem{Ash:1965}
Robert B Ash, \emph{Information Theory}. New York: Interscience, 1965. ISBN 0-470-03445-9. \\ 
New York: Dover 1990. ISBN 0-486-66521-6.

\bibitem{Eisert:2002}
Jens Eisert, Christoph Simon, Martin B. Plenio,\\
``On the quantification of entanglement in infinite-dimensional quantum systems'',\\
Journal of Physics A {\bf35} (2002) 3911--3923. doi: 10.1088/0305-4470/35/17/307


\bibitem{Yeung:2002}
Raymond W Yeung, \emph{Information Theory and Network Coding},  Springer Verlag, 2002, 2008. \\ ISBN 978-0-387-79233-0.


\bibitem{Jaynes:2003}
E T Jaynes, \emph{Probability Theory: The Logic of Science}, (Cambridge University Press, 2003). \\
 ISBN 978-0521592710

\bibitem{Watrous:2008}
John Watrous, \emph{Theory of Quantum Information}, Lecture notes, 2008. \\
{\sf https://cs.uwaterloo.ca/\~{}watrous/quant-info/lecture-notes/all-lectures.pdf}

\bibitem{area:2008}
M M Wolf, F Verstraete, M B Hastings, and J I Cirac,\\
``Area Laws in Quantum Systems: Mutual Information and Correlations'',\\
Physical Review Letters {\bf100} (2008) 070502 [arXiv: 0704.3906 [quant-ph]]. 


\bibitem{bounds:2009} 
Eric Carlen,
``Trace Inequalities and Quantum Entropy: An Introductory Course'',\\
\emph{Entropy and the Quantum: A school on analytic and functional inequalities with applications}, Tucson, Arizona, March 16-20, 2009.\\
{\sf http://www.mathphys.org/AZschool/material/AZ09-carlen.pdf}

\bibitem{Renner:2009}
Renato Renner, \emph{Quantum Information theory}, Lecture notes, 2009.\\
{\sf http://www.itp.phys.ethz.ch/education/lectures\_fs11/qit}

\bibitem{area:2010}
J Eisert, M Cramer, and M B Plenio, ``Colloquium: Area laws for the entanglement entropy'', 
Reviews of Modern Physics {\bf82} (2010) 277--306.


\bibitem{Baccetti:2012}
Valentina Baccetti and Matt Visser,
``Infinite Shannon entropy'',\\
JSTAT --- Journal of Statistical Mechanics: Theory and Experiment  {\bf 1304} (2013) P04010\\
{}[arXiv:1212.5630 [Statistical Mechanics (cond-mat.stat-mech)]]

\bibitem{Visser:2012}
Matt Visser,
``Zipf's law, power laws, and maximum entropy'',\\
New Journal of Physics  {\bf15} (2013) 043021\\ 
{}[arXiv:1212.5567 [Physics and Society (physics.soc-ph)]]. 


\bibitem{paddy1}
T. Padmanabhan, 
``Thermodynamical Aspects of Gravity: New insights'', \\
Reports in Progress of Physics {\bf73} (2010) 046901 [arXiv:0911.5004]

\bibitem{paddy2}
T. Padmanabhan, 
``Lessons from Classical Gravity about the Quantum Structure of Spacetime'', 
J.Phys. Conf. Ser. {\bf306} (2011) 012001 [arXiv:1012.4476]

\bibitem{paddy3}
T. Padmanabhan, 
``Structural Aspects Of Gravitational Dynamics And The Emergent Perspective Of Gravity'', 
AIP Conf. Proc. {\bf1483}, (2912) 2120-238

  \bibitem{Chirco:2010a}
  G.~Chirco, C.~Eling and S.~Liberati,\\
  ``The universal viscosity to entropy density ratio from entanglement'',\\
  Phys.\ Rev.\ D {\bf 82} (2010) 024010
  [arXiv:1005.0475 [hep-th]].

\bibitem{Chirco:2010b}
  G.~Chirco, C.~Eling and S.~Liberati,\\
  ``Reversible and Irreversible Spacetime Thermodynamics for General Brans-Dicke Theories'',\\
  Phys.\ Rev.\ D {\bf 83} (2011) 024032
  [arXiv:1011.1405 [gr-qc]].
  
  \bibitem{Chirco:2011}
  G.~Chirco, C.~Eling and S.~Liberati,\\
  ``Higher Curvature Gravity and the Holographic fluid dual to flat spacetime'',\\
  JHEP {\bf 1108} (2011) 009
  [arXiv:1105.4482 [hep-th]].

  
\bibitem{Unruh:1976}
W.~G.~Unruh,
  ``Notes on black hole evaporation'',
  Phys.\ Rev.\ D {\bf 14} (1976) 870.
  

  \bibitem{Abreu1} 
  G.~Abreu and M.~Visser,
  ``Tolman mass, generalized surface gravity, and entropy bounds'',\\
  Phys.\ Rev.\ Lett.\  {\bf 105}, 041302 (2010)
  [arXiv:1005.1132 [gr-qc]].

\bibitem{Abreu2} 
  G.~Abreu and M.~Visser,
  ``Entropy bounds for uncollapsed matter'',\\
  J.\ Phys.\ Conf.\ Ser.\  {\bf 314}, 012035 (2011)
  [arXiv:1011.4538 [gr-qc]].

\bibitem{Abreu3} 
G.~Abreu and M.~Visser,
  ``Entropy bounds for uncollapsed rotating bodies'',\\
  JHEP {\bf 1103}, 056 (2011)
  [arXiv:1012.2867 [gr-qc]].

\bibitem{Abreu4} 
G.~Abreu, C.~Barcel\'o and M.~Visser,
  ``Entropy bounds in terms of the $w$ parameter'',\\
  JHEP {\bf 1112}, 092 (2011)
  [arXiv:1109.2710 [gr-qc]].


  
\bibitem{twilight}
  C.~Barcel\'o and M.~Visser,
  ``Twilight for the energy conditions?'',\\
  Int.\ J.\ Mod.\ Phys.\ D {\bf 11} (2002) 1553
  [gr-qc/0205066].
 
\bibitem{B-bound}
J.~D.~Bekenstein,\\
  ``A Universal Upper Bound on the Entropy to Energy Ratio for Bounded Systems'',\\
  Phys.\ Rev.\ D {\bf 23} (1981) 287.
  
 \bibitem{stellar}
Subrahmanyan Chandrasekhar,\\
\emph{An Introduction to the Study of Stellar Structure},
(Dover, 1957). 
  

\bibitem{Barbado:2012}
  L.~C.~Barbado and M.~Visser,\\
  ``Unruh--DeWitt detector event rate for trajectories with time-dependent acceleration'',\\
  Phys.\ Rev.\ D {\bf 86} (2012) 084011
  [arXiv:1207.5525 [gr-qc]].
  
  \bibitem{Sriramkumar:1994}
  L.~Sriramkumar and T.~Padmanabhan,\\
  ``Response of finite time particle detectors in non-inertial frames and curved space-time'',\\
  Class.\ Quant.\ Grav.\  {\bf 13} (1996) 2061
  [gr-qc/9408037].
  
\bibitem{Schlicht:2003}
Sebastian Schlicht.
``Considerations on the Unruh effect: Causality and regularization'',\\
  Class.\ Quant.\ Grav.\  {\bf 21} (2004) 4647
  [gr-qc/0306022].

\bibitem{Satz:2006}
Alejandro Satz.\\
``Then again, how often does the Unruh-DeWitt detector click if we switch it carefully?'',\\
  Class.\ Quant.\ Grav.\  {\bf 24} (2007) 1719
  [gr-qc/0611067].

\bibitem{Louko:2006}
Jorma Louko and Alejandro Satz.\\
``How often does the Unruh-DeWitt detector click? Regularisation by a spatial profile'',\\
  Class.\ Quant.\ Grav.\  {\bf 23} (2006) 6321
  [gr-qc/0606067].

\bibitem{Obadia:2007}
N.~Obadia and M.~Milgrom.
``On the Unruh effect for general trajectories'',\\
  Phys.\ Rev.\ D {\bf 75} (2007) 065006
  [gr-qc/0701130 [GR-QC]].

\bibitem{Kothawala:2009}
Dawood Kothawala and T.~Padmanabhan.\\
``Response of Unruh-DeWitt detector with time-dependent acceleration'',\\
  Phys.\ Lett.\ B {\bf 690} (2010) 201
  [arXiv:0911.1017 [gr-qc]].

\bibitem{Doukas:2013} 
  J.~Doukas, S.~-Y.~Lin, B.~L.~Hu and R.~B.~Mann,
  ``Unruh Effect under Non-equilibrium conditions: Oscillatory motion of an Unruh-DeWitt detector'',\\
  JHEP {\bf 1311}, 119 (2013)
  [arXiv:1307.4360 [gr-qc]].
  
  \bibitem{Visser-nas}
  M.~Visser,
  ``Essential and inessential features of Hawking radiation'',\\
  Int.\ J.\ Mod.\ Phys.\ D {\bf 12} (2003) 649
  [hep-th/0106111].
  
  \bibitem{Barcelo-nas1}
  C.~Barcel\'o, S.~Liberati, S.~Sonego and M.~Visser,\\
  ``Hawking-like radiation does not require a trapped region'',\\
  Phys.\ Rev.\ Lett.\  {\bf 97} (2006) 171301
  [gr-qc/0607008].
  
   \bibitem{Barcelo-nas2}
  C.~Barcel\'o, S.~Liberati, S.~Sonego and M.~Visser,\\
  ``Hawking-like radiation from evolving black holes and compact horizonless objects'',\\
  JHEP {\bf 1102} (2011) 003
  [arXiv:1011.5911 [gr-qc]].
  
  \bibitem{Barcelo-nas3}
   C.~Barcel\'o, S.~Liberati, S.~Sonego and M.~Visser,\\
  ``Minimal conditions for the existence of a Hawking-like flux'',\\
  Phys.\ Rev.\ D {\bf 83} (2011) 041501
  [arXiv:1011.5593 [gr-qc]].
  
  
  \bibitem{wormholes}
  M.~Visser,
  \emph{Lorentzian wormholes: From Einstein to Hawking},\\
  (AIP Press, now Springer--Verlag, New York, 1995)
  
  \bibitem{censor}
  J.~L.~Friedman, K.~Schleich and D.~M.~Witt,
  ``Topological censorship'',
  Phys.\ Rev.\ Lett.\  {\bf 71} (1993) 1486
   [Erratum-ibid.\  {\bf 75} (1995) 1872]
  [gr-qc/9305017].
   
  \bibitem{Flanagan:1996}
  E.~E.~Flanagan and R.~M.~Wald,\\
  ``Does back reaction enforce the averaged null energy condition in semiclassical gravity?'',\\
  Phys.\ Rev.\ D {\bf 54} (1996) 6233
  [gr-qc/9602052].
    
  \bibitem{q-anec}
  C.~J.~Fewster and T.~A.~Roman,
  ``Null energy conditions in quantum field theory'',\\
  Phys.\ Rev.\  D {\bf 67} (2003) 044003
  [arXiv:gr-qc/0209036].
  

\bibitem{Ford:1990}
  L.~H.~Ford,
  ``Constraints on negative energy fluxes'',
  Phys.\ Rev.\  D {\bf 43}, 3972 (1991).
  
  \bibitem{Ford+Roman}
  L.~H.~Ford and T.~A.~Roman,
  ``Averaged Energy Conditions And Quantum Inequalities'',\\
  Phys.\ Rev.\  D {\bf 51} (1995) 4277
  [arXiv:gr-qc/9410043].
  
  
  \bibitem{Ford:1996}
  L.~H.~Ford and T.~A.~Roman,
  ``Restrictions on negative energy density in flat spacetime'',\\
  Phys.\ Rev.\  D {\bf 55}, 2082 (1997)
  [arXiv:gr-qc/9607003].

\bibitem{Ford+Roman+collaborators}
L.~H.~Ford, M.~J.~Pfenning and T.~A.~Roman,\\
  ``Quantum inequalities and singular negative energy densities'',\\
  Phys.\ Rev.\  D {\bf 57} (1998) 4839
  [arXiv:gr-qc/9711030].

\bibitem{Ford:1999}
  L.~H.~Ford and T.~A.~Roman,
  ``The quantum interest conjecture'',\\
  Phys.\ Rev.\  D {\bf 60}, 104018 (1999)
  [arXiv:gr-qc/9901074].
  

\bibitem{Fewster:1999a}
  C.~J.~Fewster and E.~Teo,
  ``Quantum inequalities and `quantum interest' as eigenvalue problems'',\\
  Phys.\ Rev.\  D {\bf 61}, 084012 (2000)
  [arXiv:gr-qc/9908073].
  
  \bibitem{Fewster:1999b}
  C.~J.~Fewster,
  ``A General worldline quantum inequality'',\\
  Class.\ Quant.\ Grav.\  {\bf 17} (2000) 1897
  [gr-qc/9910060].

  \bibitem{Fewster:2007}
  C.~J.~Fewster and C.~J.~Smith,
  ``Absolute quantum energy inequalities in curved spacetime'',\\
  Annales Henri Poincare {\bf 9} (2008) 425
  [gr-qc/0702056 [GR-QC]].


\bibitem{Prado1}
P.~Mart\'in-Moruno and M.~Visser,\\
  ``Classical and quantum flux energy conditions for quantum vacuum states'',\\
  Physical Review {\bf D88} (2013) 061701R [arXiv:1305.1993 [gr-qc]].

\bibitem{Prado2}
  P.~Mart\'in-Moruno and M.~Visser,
  ``Semiclassical energy conditions for quantum vacuum states'',\\
  JHEP {\bf1309} (2013) 050 [arXiv:1306.2076 [gr-qc]].
  

  
  
\end{thebibliography}
\end{document}